\documentclass[11pt,a4paper]{article}
\usepackage{jheppub}
\usepackage{amsmath}
\usepackage{amssymb}
\usepackage{xcolor}
\usepackage{float}
\usepackage{enumerate}
\usepackage{slashed}

\newcommand{\bea}{\begin{eqnarray}}
\newcommand{\eea}{\end{eqnarray}}
\newcommand{\bit}{\begin{itemize}}
\newcommand{\eit}{\end{itemize}}

\def\nl{\nonumber \\}

\def\a{\alpha}

\def\b{\beta}

\def\s{\sigma}

\def\p{\partial}

\def\le{\left(}
\def\ri{\right)}

\def\beq{\begin{equation}}
\def\eeq{\end{equation}}

\def\arr{{\rightarrow}}

\title{Trace anomaly for non-relativistic fermions} 

\author[a,b]{Roberto Auzzi,} 
\author[c]{Stefano Baiguera} 
\author[a,d]{and Giuseppe Nardelli}

\affiliation[a]{Dipartimento di Matematica e Fisica, Universit\`a Cattolica
del Sacro Cuore, \\
Via Musei 41, 25121 Brescia, Italy}
\affiliation[b]{INFN Sezione di Perugia, \\ Via A. Pascoli, 06123 Perugia, Italy}
\affiliation[c]{Universit\`a degli studi di Milano Bicocca and INFN, Sezione di Milano - Bicocca, \\ Piazza
della Scienza 3, 20161, Milano, Italy}
\affiliation[d]{TIFPA - INFN, c/o Dipartimento di Fisica, Universit\`a di Trento, 38123 Povo (TN), Italy}

\emailAdd{roberto.auzzi@unicatt.it}
\emailAdd{giuseppe.nardelli@unicatt.it}
\emailAdd{s.baiguera@campus.unimib.it}

\abstract{
We study the coupling of a $2+1$ dimensional  non-relativistic  spin $1/2$
fermion to a curved 
Newton-Cartan geometry, using null reduction  from an 
extra-dimensional relativistic Dirac action in curved spacetime.
We analyze Weyl invariance in detail: we show that at the classical level
it is preserved in an arbitrary curved background, whereas at the quantum
level it is broken by anomalies.
We compute the trace anomaly using the Heat Kernel method
and we show that the anomaly coefficients $a$, $c$
are proportional to the relativistic ones for a Dirac fermion in $3+1$ dimensions.
As for the previously studied scalar case, these coefficents
 are proportional to $1/m$, where $m$ is the non-relativistic
 mass of the particle.
}

\keywords{}

\begin{document}

\maketitle

\section{Introduction}

Newton-Cartan (NC) geometry was originally proposed as
a covariant formulation of Newtonian gravity 
(see e.g. \cite{gravitation} for a review).
In recent times it raised growing interest for applications to condensed matter
systems (see e.g. \cite{Son:2005rv,Hoyos:2011ez,Son:2013rqa,Geracie:2014nka}), 
such as fermions at unitarity and quantum Hall effect.
The background fields of NC gravity provide a natural set of sources
for operators in the energy-momentum tensor multiplet of theories
with non-relativistic  Schr\"odinger invariance.

Many theoretical difficulties in dealing with these systems are due
to the strong coupling nature of the interaction. Strong coupling may
drastically change the infrared (IR) degrees of freedom
coming from a given ultraviolet (UV) description. 
Renormalization Group (RG) trajectories may interpolate from weak to strong coupling changing
the nature of the physical spectrum and of the degrees of freedom.
In relativistic theories there are general results which
formalize the intuition that information is lost
when coarse graining is implemented from UV to IR, namely
Zamolodchikov's $c$-theorem in $d=2$ \cite{Zamolodchikov:1986gt},
 the $F$-theorem in $d=3$  \cite{Jafferis:2010un,Jafferis:2011zi,Pufu:2016zxm}
  the $a$-theorem in $d=4$ 
  \cite{Cardy:1988cwa,Osborn:1989td,Jack:1990eb,Osborn:1991gm,Komargodski:2011vj,Komargodski:2011xv}. 
For condensed matter applications, it would be interesting 
to establish similar results in  non-relativistic systems.

With these motivations, in the last years a certain amount of work
has been  devoted to the study of non-relativistic trace anomalies.
In general, trace anomalies can be classified into two classes \cite{Deser:1993yx}:
type A or B depending if they have non-vanishing or vanishing Weyl variation, respectively.
The relevant ones for RG constraints are the type A,
such as $c$ in $d=2$ or $a$ in $d=4$ relativistic systems.

In the non-relativistic case,
at a fixed point space and time may have different relative scaling,
which can be parameterized by the dynamical exponent $z$:
\beq
x^i \arr e^\s x^i \, , \qquad t \arr e^{z \s} t \, .
\eeq
 Moreover one may distinguish between  Schr\"odinger and Lifshitz systems,
whose main difference relies on the presence of Galilean boost invariace.
So far, in all the known cases, the Lifshitz trace anomalies
(see \cite{Adam:2009gq,Baggio:2011ha,Griffin:2011xs,Arav:2014goa,Arav:2016xjc,Barvinsky:2017mal,Arav:2016akx})
 turn out
to be of type B and so they do not give interesting candidates for 
monotonic quantities.
 In the Schr\"odinger case, 
 in $d=2+1$ dimensions and for dynamical exponent $z=2$,
 if one couples the theory to a 
curved NC background, it exists a type A anomaly 
 \cite{Jensen:2014hqa}\footnote{
An interesting subtlety is that, in order to get a type-A anomaly,
gravity backgrounds which do not satisfy Frobenius condition
must be considered \cite{Auzzi:2015fgg,Arav:2016xjc}.}. 
The structure of this anomaly is the same as the trace anomaly
for $d=4$ relativistic theories, and so
it includes a type-$A$ and a type-$B$ part, parameterized by  $a$
and  $c$ coefficients:
\beq
\mathcal{A}=
 -a \, E_4 + c \, W^2 + \dots \, .
\eeq
In this equation $E_4$ and $W^2$ are the Euler density
and the Weyl tensor squared of the null reduction metric in eq.~(\ref{null-red});
these quantities are completely determined in terms of $2+1$-dimensional
NC geometry data. The use of the extra dimension is a formidable trick
to conveniently keep track of the Milne boost symmetry. 
Cohomological analysis and general properties were studied in 
 \cite{Jensen:2014hqa,Auzzi:2015fgg,Arav:2016xjc,Auzzi:2016lrq}.
The first explicit calculation of anomalies for a physical system
was performed in \cite{Auzzi:2016lxb} with the Heat Kernel (HK) method, for the case of a free scalar.
Later this result was confirmed in \cite{Fernandes:2017nvx}  using Fujikawa approach\footnote{
In \cite{Fernandes:2017nvx}  an extra term is present, which could not be detected with the background
chosen in \cite{Auzzi:2016lxb};
this issue deserves further study which goes beyond the purpose of the present paper.
 On the other hand, both the calculations in \cite{Fernandes:2017nvx,Auzzi:2016lxb} 
disagree with \cite{Pal:2017ntk}.}.

Fermions are a fundamental ingredient in Nature; the purpose
of this paper is to study conformal invariance and anomalies
for a free non-relativistic spin $1/2$ fermion coupled to a generic 
curved Newton-Cartan background, using  
null reduction from a $3+1$ dimensional
relativistic action.
First of all, we show that it is possible
to couple the fermion to the geometry in a Weyl invariant way;
this is not trivial, due to the different scaling properties of 
the components of frame fields, spin connection and
dynamical fermionic fields.
 Our analysis specializes to the case 
where the gyromagnetic ratio $g$ is twice the spin $s$;
the generic case requires modified Milne boost 
transformations \cite{Geracie:2014nka}
on the sources and can not be studied by null reduction.

The other issue that we address is the computation of the anomaly 
coefficients $a$ and $c$ using Heat Kernel. In the bosonic case, these coefficients
turn out to be proportional to the corresponding ones 
in relativistic systems in $3+1$ dimensions. We find that the same property
still persists also in the fermionic case.

The paper is organized as follows. In sect. \ref{sect-null-reduction} we derive the fermionic action
from the null reduction of the Dirac one and we discuss the gyromagnetic factor.
In sect. \ref{sect-weyl} we show in detail that the action is Weyl invariant.
In sect. \ref{sect-heat} we compute the trace anomaly using HK method.
We conclude in sect. \ref{sect-concl}, tecnical details are in appendices.

\section{Null reduction for fermions}
\label{sect-null-reduction}

\subsection{Metric and frame fields}

We will consider the coupling of non-relativistic fermions
in $2+1$ dimensions to a background NC
geometry.  In order to make the implementation of the local version of the Galilean 
symmetry (Milne boost invariance) more covenient, we use the null-reduction 
method \cite{Duval:1984cj} from an extra-dimensional relativistic 
$3+1$ dimensional theory. We will sometimes refer to null-reduction method
as Discrete Light-Cone Quantization (DLCQ).
Useful references about NC geometry include 
\cite{Jensen:2014aia,Hartong:2014pma,Hartong:2014oma,Hartong:2015wxa,
Geracie:2015dea,Geracie:2015xfa,Geracie:2016dpu,Geracie:2016bkg}.  
Galilei invariance for fermions was first studied
in \cite{LevyLeblond:1967zz}.
For other approaches to couple non-relativistic theories 
to background NC geometry see \cite{Geracie:2014nka,Fuini:2015yva}.
Other applications of null reduction to fermions were
discussed in \cite{Duval:1995fa}.

In our conventions late latin capital  indices,  like $ M, N, \ldots $,  correspond to  $3+1$ dimensional curved space-time indices, whereas early latin capital indices  like $A, B, \ldots , $ correspond to tangent space indices, where the metric is locally flat. 
The coordinate  $x^-$ denotes the null direction 
of the dimensional reduction.
The remaining light-cone coordinate, 
$x^+$, will play the role of time in the lower dimensional non-relativistic theory. Curved space coordinates will be labelled by lower case latin indices $i,j, \ldots$, whereas the tangent space counterparts will be labelled by $a,b,\ldots$. Collectively, space time indices of the lower dimensional theory will be denoted by $\mu , \nu , \ldots$ and $\alpha, \beta , \ldots$ for curved and tangent space coordinates, respectively.
Summarizing,  DLCQ indices are
\beq
\begin{aligned}
&  M=(-,\mu)=(-,+,i)    \qquad (i=1,2)  \\
&  A=(-,\alpha)=(-,+,a)  \qquad  (a=1,2)  \, .
\end{aligned}
\eeq
Since the light-cone indices $\pm$ use the same symbols for  curved or tangent space indices,  we will use the notations 
\beq
\underset{(M)}{\pm} \, , \qquad \underset{(A)}{\pm}  \, ,
\eeq
indicating that they  refer to curved (subscript $(M)$) or tangent space  (subscript $(A)$) light-cone coordinates.

In order to apply the null reduction, we will consider fields of the form
\beq
\Psi(x^M)=\psi(x^{\mu}) e^{imx^{-}}
\eeq
and a metric of the form
\beq
\label{null-red}
G_{MN}= \begin{pmatrix}
0  & n_{\nu}  \\
n_{\mu}  & n_{\mu} A_{\nu} + n_{\nu} A_{\mu} + h_{\mu\nu}
\end{pmatrix} \, ,  \qquad
G^{MN}= \begin{pmatrix}
A^2 - 2 v \cdot A  & v^{\nu}-h^{\nu\sigma} A_{\sigma}  \\
v^{\mu}-h^{\mu\sigma} A_{\sigma}  & h^{\mu\nu}
\end{pmatrix} \, .
\eeq
We denote the determinant of the metric as:
\beq
\sqrt{g}= \sqrt{- \det G_{AB}}=
\sqrt{{\rm det}(h_{\mu \nu}+ n_\mu n_\nu)}
\eeq

The metric tensor $G_{MN}$ defines a non degenerate $3+1$ dimensional metric whose entrees encode the main ingredients of the $2+1$ dimensional NC geometry: a positive definite symmetric rank $2$ tensor $h^{\mu \nu}$, which corresponds to the spatial inverse metric, and a nowhere-vanishing vector $n_\mu$ (defining the local time direction),
with the condition $
n_\mu h^{\mu \a}=0$. 
In order to define a spatial metric with lower indices and a connection, one introduces
 a velocity field $v^\mu$, with the condition $n_\mu v^\mu =1$.
 Given 
$(h^{\mu \nu}, n_\mu, v^\nu)$,  one can then uniquely define $h_{\mu \nu}$,
with:
\beq
h^{\mu \rho}h_{\rho \nu}=\delta^\mu_\nu -v^\mu n_\nu\equiv P^\mu_\nu \, , \qquad
h_{\mu \a} v^\a =0 \, ,
 \eeq
 where $P^\mu_\nu$ is the projector onto spatial directions. The velocity vector is not unique (it is only required to satisfy 
 $n_\mu v^\mu=1$) and the ambiguity in the choice of $v$ is related to the last ingredient of the NC geometry: a non-dynamical gauge field $A_\mu$,
 whose presence is necessary to guarantee Milne boost invariance. This gauge field will act as a source for the particle number symmetry. 
We introduce, for later convenience,  the antisymmetric tensors:
\beq
\tilde{F}_{\mu\nu} = \p_{\mu} n_{\nu} - \p_{\nu} n_{\mu} \, , 
\qquad
F_{\mu\nu} = \p_{\mu} A_{\nu} - \p_{\nu} A_{\mu} \, .
\eeq

The null reduction is a useful trick to realise the invariance
under the following Milne boost transformations:
\bea
v'^\mu & = & v^\mu+h^{\mu \nu} \psi_\nu \, \nl
h'_{\mu \nu} & = & h_{\mu \nu} -(n_\mu P_\nu^\rho+ n_\nu P_\mu^\rho) \psi_\rho
+n_\mu n_\nu h^{\rho \sigma} \psi_\rho \psi_\sigma \, , \nl
A'_\mu & = & A_\mu+P^\rho_\mu \psi_\rho -\frac{1}{2} n_\mu h^{\a \b} \psi_\a \psi_\b \, , 
\eea
while $n_\mu$ and $h^{\mu \nu}$ are invariant.
Modified Milne transformation may also be considered, 
but then the null reduction trick can not be used (see e.g. \cite{Jensen:2014aia}).

Since we are dealing with spinors, the covariant derivative also contains the spin connection term and then it is necessary to introduce an orthonormal frame field (vielbein) which relates the metric in the curved spacetime with the flat tangent space.
The metric in the flat tangent space is given by
\beq
G_{AB}= G^{AB} = \begin{pmatrix}
0  & 1 & 0 & 0  \\
1  & 0 & 0 & 0  \\
0  & 0 & 1 & 0  \\
0  & 0 & 0 & 1  \\
\end{pmatrix} \, . 
\eeq
As usual, the vielbein are defined by the following relations:
\bea
& G_{MN}= e^{A}_{\,\,\, M} G_{AB} e^{B}_{\,\,\, N}  \, ,  
\qquad  G_{AB}= e^{M}_{\,\,\, A} G_{MN} e^{N}_{\,\,\, B}  \, ,  
\nl
&  e^{A}_{\,\,\, M} e^{M}_{\,\,\, B} = \delta^{A}_{\,\,\, B}  \, , \qquad
  e^{M}_{\,\,\, A} e^{A}_{\,\,\, N} = \delta^{M}_{\,\,\, N}  \, .
\eea

In order to consider the coupling of fermions to $2+1$ NC
gravity, the dreibein will be defined by dimensional reduction of fierbein. Such operation is not unique.
The following choice 
turns out to be convenient:
\beq
e^A_{\,\,\, M} = \begin{pmatrix}
e^-_{\,\,\, M}  \\
e^+_{\,\,\, M}  \\
e^a_{\,\,\, M}
\end{pmatrix} = 
 \begin{pmatrix}
e^-_{\,\,\, -} \ \ & e^-_{\,\,\, \mu}  \\
e^+_{\,\,\, -} \ \ & e^+_{\,\,\, \mu} \\
e^a_{\,\,\, -} \ \ & e^a_{\,\,\, \mu}
\end{pmatrix}
=
 \begin{pmatrix}
1 \ \ & A_{\mu} \\
0 \ \ & n_{\mu} \\
\mathbf{0} \ \ & e^a_{\mu}
\end{pmatrix}
\, .
\eeq

We can  further simplify our expression by using the consistency relations among fielbein with different indices
\beq
e^{M}_{\,\,\, A} e^{B}_{\,\,\, M} = \delta_{A}^{\,\,\,B}  \, , \qquad  
e^{A}_{\,\,\, M} e^{N}_{\,\,\, A} = \delta_{M}^{\,\,\,N}  \, ,
\label{4 consistenza tra vielbein}
\eeq
which entail the following constraints:
\beq
\begin{aligned}
& v^{\mu} e^a_{\,\,\, \mu} = 0  \qquad & (\textrm{from} \,\,\, e^M_{\,\,\, +} e^{a}_{\,\,\, M}=0)\\
& n_{\mu} e^{\mu}_{\,\,\, a} = 0  \qquad & (\textrm{from} \,\,\, e^M_{\,\,\, a} e^{+}_{\,\,\, M}=0) \\
&   e^{\mu}_{\,\,\, a} e^b_{\,\,\, \nu} = \delta_a^{\,\,\, b}  \qquad & (\textrm{from} \,\,\, e^M_{\,\,\, a} e^{b}_{\,\,\, M}=0)  \\
& h^{\mu\nu}   e^a_{\,\,\, \mu}  e^b_{\,\,\, \nu} = \delta^{ab}  \qquad & (\textrm{from} \,\,\, e^M_{\,\,\, a} e^{b}_{\,\,\, M}=0) \\
&  e^a_{\,\,\, \mu} e^a_{\,\,\, \nu} = h_{\mu\nu} \, , \qquad   \qquad e^{\mu}_{\,\,\, a} e^{\nu}_{\,\,\, a} =h^{\mu\nu} \, ,
 & e^{\mu}_{\,\,\, a} = e^{\mu a}  \\
\end{aligned}
\eeq
These relations simplify the vielbein with the inverted indices:
\beq
e^M_{\,\,\, A} = \begin{pmatrix} e^M_{\,\,\, -}\ \ & e^M_{\,\,\, +} \ \ & e^M_{\,\,\, a}   \end{pmatrix} =
 \begin{pmatrix} e^-_{\,\,\, -}\ \ & e^-_{\,\,\, +} \ \ & e^-_{\,\,\, a}  \\
 e^{\mu}_{\,\,\, -}\ \ & e^{\mu}_{\,\,\, +} \ \ & e^{\mu}_{\,\,\, a}  \end{pmatrix}
= \begin{pmatrix}1 \ \ & -v^{\sigma} A_{\sigma} \ \ &  - h^{\nu \sigma} A_{\sigma} e^a_{\nu}  \\
\bold{0} \ \ &  v^{\mu} \ \ &   h^{\mu\nu} e^{a}_{\nu} \end{pmatrix} \, .
\eeq
The following relation is useful:
\beq
h_{\mu \rho} e^\rho_a=h_{\mu \rho} h^{\rho \tau} e^a_\tau=(\delta_\mu^\tau -v^\tau n_\mu) e^a_\tau= e^a_\mu \, .
\eeq


\subsection{Dirac action}

The  Dirac operator is expressed as
\beq
\slashed D= \gamma^{M} D_{M} = \gamma^{A} e^{M}_{\,\,\, A} D_{M}  \, ,
\eeq
Conventions for gamma matrices with lightcone indices are
summarized in appendix \ref{app-gamma}.
The covariant derivative takes the form 
\beq
D_{M} \Psi =\left( \p_{M} + \frac{1}{4} \omega_{MAB} \gamma^{AB} \right) \Psi = \left( \p_{M} + \frac{1}{8} \omega_{MAB} [\gamma^{A}, \gamma^{B}] \right) \Psi \, ,
\eeq
$\omega_{MAB}$ being the spin connection defined in Appendix \ref{spinc}.
We shall derive the non-relativistic fermion action in $2+1$-dimensions
from the null reduction of the $3+1$-dimensional Dirac action:
\beq
S= \int d^4 x \sqrt{g} \, i \bar{\Psi} \slashed D \Psi \, .
\label{didirac}
\eeq
The connection in the covariant derivative $D_M$ 
has no torsion term and so the lagrangian 
in eq.~(\ref{didirac}) can be made hermitian 
by partial integration.


\subsection{Flat space-time}

We start by considering the simplest flat case:
\beq
n_{\mu} = (1,0,0 ) \, , \quad  h_{\mu\nu} = {\rm diag}(0,1,1) \, ,
 \quad v^{\mu} = (1,0,0) \, ,
\label{flatland}
\eeq
and $A_\mu=0$.
The Dirac action is just the flat one.
The following notation is used:
\beq
\Psi(x^{M}) = \begin{pmatrix}
\chi_L (x^{\mu})  \\
\varphi_L (x^{\mu}) \\
\varphi_R (x^{\mu})\\
\chi_R (x^{\mu})
\end{pmatrix} e^{imx^{-}}  \, .
\eeq
and the Dirac Lagrangian can be written as
\beq
\begin{aligned}
\mathcal{L} & =  - \sqrt{2} m \chi_L^{\dagger} \chi_L 
- \sqrt{2} m \chi_R ^{\dagger} \chi_R  
- \sqrt{2}i \varphi_L^{\dagger} \p_{t} \varphi_L 
- \sqrt{2}i \varphi_R^{\dagger} \p_{t} \varphi_R + \\
& +  i \varphi_L^{\dagger} (\p_1 + i \p_2) \chi_L 
+  i \chi_L^{\dagger} (\p_1 - i \p_2) \varphi_L 
- i  \chi_R ^{\dagger}  (\p_1 + i \p_2) \varphi_R 
-  i \varphi_R^{\dagger} (\p_1 - i \p_2) \chi_R    \, .
\end{aligned} 
\eeq
We find the Euler-Lagrange equations of motion for the various components:
\bea
\chi_L &=& \frac{i}{m} \frac{1}{\sqrt{2}} (\p_1 - i \p_2)  \varphi_L \, , \qquad
\chi_R = - \frac{i}{m}  \frac{1}{\sqrt{2}} (\p_1 + i \p_2) \varphi_R  \, , 
\nl
\p_t \varphi_L &=&    \frac{1}{\sqrt{2}} (\p_1 + i \p_2) \chi_L  \, , \qquad
  \p_t \varphi_R   = - \frac{1}{\sqrt{2}} (\p_1 - i \p_2)  \chi_R    \, .
 \label{eq-free}
\eea
As expected for the Dirac action in the massless case, 
the left and right Weyl spinors decouple.
The auxiliary fields $\chi_{L,R}$ can be eliminated
by the equations of motion and replaced
in the Lagrangian;
 we obtain a set of  
decoupled Schr\"odinger 
 equations for the fermions $\varphi_{L,R}$.


\subsection{Curved spacetime}

In this section we will write eq.~(\ref{didirac}) in a more explicit way,
in order  to later 
establish the gyromagnetic factor
and show that the action is conformal invariant.
The left and right-handed parts of the Dirac spinor decouple:
\beq
\Psi=  \begin{pmatrix}
\Psi_L \\ \Psi_R 
\end{pmatrix} \, , \qquad 
\Psi_L=  \begin{pmatrix}
\chi_L \\ \varphi_L 
\end{pmatrix} \, , \qquad
\Psi_R=  \begin{pmatrix}
 \varphi_R \\
 \chi_R 
\end{pmatrix} \, .
\eeq
In the remaining part of this section we will consider
the action $\mathcal{L}_1$ for just the left component $\Psi_L$
($\Psi_R$ is completely analogous): 
\beq
\Psi_L=\begin{pmatrix}
\chi \\ \varphi 
\end{pmatrix}  e^{im x^-}\,
\, , \qquad
S_{L}=\int d^4 x \sqrt{g} \,  \mathcal{L}_1 \, ,
\eeq
where
\beq
\begin{aligned}
\mathcal{L}_1  = i \Psi_L^{\dagger} \bar{\sigma}^A D_A \Psi_L & =
i e^{-imx^-} \begin{pmatrix}  \chi^{\dagger} & \varphi^{\dagger}  \end{pmatrix} \bar{\sigma}^-  D_{\underset{(A)}{-}} \left[  \begin{pmatrix}  \chi \\ \varphi  \end{pmatrix}  e^{imx^-}  \right] + \\
& + i e^{-imx^-} \begin{pmatrix}  \chi^{\dagger} & \varphi^{\dagger}  \end{pmatrix}  \bar{\sigma}^+ D_{\underset{(A)}{+}} \left[  \begin{pmatrix}  \chi \\ \varphi  \end{pmatrix}  e^{imx^-}  \right] + \\
& + i e^{-imx^-} \begin{pmatrix}  \chi^{\dagger} & \varphi^{\dagger}  \end{pmatrix} \sigma^a  e^M_{\,\,\, a}  D_{M} \left[  \begin{pmatrix}  \chi \\ \varphi  \end{pmatrix}  e^{imx^-}  \right] \, .
\end{aligned}
\eeq
For convenience, we renamed $(\chi_L,\varphi_L)$ as $(\chi, \varphi)$.
An explicit calculation gives:
\beq
\begin{aligned}
 D_{\underset{(A)}{-}} = e^M_{\,\,\, \underset{(A)}{-}} D_M = 
\begin{pmatrix} 1 & \bold{0} \end{pmatrix} \begin{pmatrix} D_{\underset{(M)}{-}} \\ D_{\mu} \end{pmatrix} = D_{\underset{(M)}{-}} \, ,
\\
 D_{\underset{(A)}{+}} = e^M_{\,\,\, \underset{(A)}{+}} D_M = 
\begin{pmatrix} -v^{\sigma} A_{\sigma} & v^{\mu} \end{pmatrix} \begin{pmatrix} D_{\underset{(M)}{-}} \\ D_{\mu} \end{pmatrix} = - v^{\sigma} A_{\sigma} D_{\underset{(M)}{-}} + v^{\mu} D_{\mu}  \, ,
\\
 D_{a} = e^M_{\,\,\, a} D_M = 
\begin{pmatrix} -e^{\sigma}_{\,\,\, a} A_{\sigma} & e^{\mu}_{\,\,\, a} \end{pmatrix} \begin{pmatrix} D_{\underset{(M)}{-}} \\ D_{\mu} \end{pmatrix} =  -e^{\sigma}_{\,\,\, a} A_{\sigma}  D_{\underset{(M)}{-}} + e^{\mu}_{\,\,\, a} D_{\mu}  \, .
\end{aligned}
\eeq
We can write $\mathcal{L}_1$ as follows:
\beq
 \begin{aligned}
\mathcal{L}_1 & = - \sqrt{2} m \chi^{\dagger} \chi - \sqrt{2} i \varphi^{\dagger} \hat{D}_{t} \varphi + i \varphi^{\dagger} (\hat{D}_{1} +i \hat{D}_2 ) \chi + i \chi^{\dagger}(\hat{D}_{1} - i \hat{D}_2 ) \varphi + \\
& +  \frac{i}{4} \begin{pmatrix}  \chi^{\dagger} & \varphi^{\dagger}  \end{pmatrix} 
(\bar{\sigma}^+  v^{\mu} +\sigma^a  e^{\mu}_{\,\,\, a} )
\omega_{\mu AB} \sigma^{AB}  \begin{pmatrix}  \chi \\ \varphi  \end{pmatrix}  \\
& +  \frac{i}{4}  \begin{pmatrix}  \chi^{\dagger} & \varphi^{\dagger}  \end{pmatrix} \left( \bar{\sigma}^-  - v^{\sigma} A_{\sigma} \bar{\sigma}^+ - \sigma^a e^{\sigma}_{\,\,\, a} A_{\sigma}  \right) \omega_{ {\underset{(M)}{-}} AB} \sigma^{AB}   \begin{pmatrix}  \chi \\ \varphi  \end{pmatrix}  \, , 
\end{aligned} 
\label{Lagrangiana-ferm}
\eeq
where we introduced derivatives which are covariant with respect to the local U(1) symmetry:
\beq
\hat{D}_t = v^{\mu} \left( \p_{\mu} -i m A_{\mu}  \right) \, , \qquad
\hat{D}_a = e^{\mu}_{\,\,\, a} \left( \p_{\mu} -i m A_{\mu}  \right) \, . 
\eeq
The last two lines are more troublesome and require the explicit knowledge of the components of the spin connection, because its Lorentz indices are contracted with sigma matrices, containing also spinorial indices.

We can put the  action (\ref{Lagrangiana-ferm}) in the following form:
\beq
\mathcal{L}_1=
 \begin{pmatrix}  \chi^{\dagger} & \varphi^{\dagger}  \end{pmatrix} 
  \begin{pmatrix} A  & B \\ C & E \end{pmatrix} 
 \begin{pmatrix}  \chi \\ \varphi  \end{pmatrix} \, , 
 \label{Lagrangiana-ferm2}
\eeq
where
\bea
A &=& -\sqrt{2}\le m+\frac14 \tilde{F}_{\mu \nu} e^\mu_1 e^\nu_2\ri\, ,
\nl
B &=& (e_1^\mu-i e^\mu_2) (i \tilde{D}_\mu +\frac{i}{4} \tilde{F}_{\mu \nu} v^\nu) \, ,
 \qquad
C =(e_1^\mu+i e^\mu_2) (i \tilde{D}_\mu +i \frac{ 3}{4} \tilde{F}_{\mu \nu} v^\nu) \,  ,
\nl
E &=& \sqrt{2} \left[  v^\mu (-i \tilde{D}_\mu 
-\frac{i}{4} h^{\rho \sigma} \p_\mu h_{\rho \sigma})
-\frac{i}{2} (v^\mu v^\nu \p_\mu n_\nu +\p_\mu v^\mu)
-\frac14 F_{\mu \nu}  e^\mu_1 e^\nu_2  
\right] \, .
\eea
In these expressions
$\tilde{D}_\mu$ 
denotes a partially covariant derivative
which includes just the gauge and the curved space spin
connection  $\tilde{\omega}_{\mu  a b}$ built just 
with the spatial tetrad $e^a_\mu$; this derivative acts
on the matter fields $\varphi$ and $\chi$ as follows:
\beq
\tilde{D}_\mu \varphi=
\left[ \p_\mu -\frac{i}{2} \tilde{\omega}_{\mu 1 2} 
- i m A_\mu \right] \varphi  \, ,  \qquad
\tilde{D}_\mu \chi=
\left[ \p_\mu + \frac{i}{2} \tilde{\omega}_{\mu 1 2} 
- i m A_\mu \right] \chi  \, ,  
\eeq
where
\beq
\tilde{\omega}_{\mu  a b}  = \frac{1}{2}
\le e^{\nu}_{\,\,\,a} \left(  \p_{\mu} e^{b}_{\,\,\, \nu} -  \p_{\nu} e^{b}_{\,\,\, \mu}  \right)  -   e^{\nu}_{\,\,\, b} \left(  \p_{\mu} e^{a}_{\,\,\, \nu} -  \p_{\nu} e^{a}_{\,\,\, \mu}   \right)  
 -e^{\nu}_{\,\,\,a} e^{\rho}_{\,\,\, b}    e^c_{\,\,\, \mu} \left( \p_{\nu} e^{c}_{\,\,\, \rho}-
   \p_{\rho} e^{c}_{\,\,\, \nu}  \right) \ri \, . 
\eeq
The auxiliary field $\chi$ is determined by the equations of motion as follows:
\beq
\chi=\frac{i(e^\mu_1-i e^\mu_2)\le \tilde{D}_\mu+\frac14 v^\nu \tilde{F}_{\mu \nu}\ri \varphi}
{\sqrt{2} \le m+\frac{\tilde{F}_{\mu \nu} e_1^\mu e_2^\nu}{4}\ri}
\label{chichi} \, .
\eeq
Replacing it into the action in eq.~(\ref{Lagrangiana-ferm2}),
we could obtain a cumbersome
Lagrangian written only in terms of $\varphi$.
 In order to keep  our calculations simple,
 we will later specialize to some specific backgrounds.

\subsection{Gyromagnetic ratio}

Let us compute the gyro-magnetic ratio of 
the non-relativistic fermion. We consider 
flat background space-time
as in eq.~(\ref{flatland})
 and generic gauge field $A_\mu$.
Specializing the general results in Appendix \ref{spinc},
we find the non-zero components of the spin connection:
\beq
\omega_{++i}=-F_{0 i}=-E_i \, , \qquad
\omega_{i+j}=-\frac{1}{2} F_{ij} = -\frac{B}{2} \, , \qquad
\omega_{0ij}=-\frac12 F_{ij} = -\frac{B}{2} \, .
\eeq
Eliminating $\chi$ with the equations of motion, we obtain:
\beq
S = \int d^3 x \, \left[ \frac{i}{2} \varphi^{\dagger} \overset{\leftrightarrow}{\p_t} \varphi - \frac{1}{2m} \delta^{ij} (D_i \varphi)^{\dagger} (D_j \varphi)  -  \frac{1}{4}B \varphi^{\dagger} \varphi  \right] \, .  
\eeq
Since the charge associated to the magnetic field is \emph{m}, we find a coupling to the magnetic field with gyromagnetic ratio $g=1$:
\beq
g \frac{q}{2 m} \vec{S} \cdot \vec{B}
\eeq
where in our case the charge $q=m$ and $\vec{S}=\vec{\sigma}/2$ is the spin. 
This is consistent with the form of the Milne boost
transformations which come from null reduction,
which are valid for $g=2s$ \cite{Geracie:2014nka}.


\section{Weyl invariance}
\label{sect-weyl}

In order to study the conformal symmetry of the theory, 
it is useful to determine the Weyl weights of the fields appearing in the action. 
Weyl transformations act on the metric in  the following way:
\beq
n_\mu \arr e^{2 \s} n_\mu \, , \qquad
h_{\mu \nu} \arr e^{2 \s} h_{\mu \nu} \qquad
v^\mu \arr e^{-2 \s} v^\mu \, , \qquad
h^{\mu \nu} \arr e^{-2 \s} h^{\mu \nu} \, .
\eeq
The action on the frame fields is as follows:
\bea
e^-_M &\arr& e^-_M \, , \qquad
e^+_M \arr e^+_M e^{2 \sigma} \, , \qquad
e^a_M \arr e^a_M e^{ \sigma}
\nl
e^M_- &\arr& e^M_- \, , \qquad
e^M_+ \arr e^M_+ e^{-2 \sigma} \, , \qquad
e^M_a \arr e^M_a e^{ -\sigma} \, .
\eea
It is also useful to know how each element of the spin connection 
transforms under a Weyl transformation:
\bea
\label{weyl-spinc}
\omega_{-ab} & \arr &  \omega_{-ab} \, , \qquad 
 \omega_{-+ a}  \arr e^{-\s} ( \omega_{-+ a} + e^\nu_a \p_\nu \s ) \, , \qquad
 \omega_{\mu-a} \arr e^\s (  \omega_{\mu-a}+ n_\mu e^\nu_a \p_\nu \s) \, ,
\nl
 \omega_{\mu - + } & \arr & \omega_{\mu - + } -\p_\mu \s +n_\mu v^\nu \p_\nu \s \, ,
\qquad
\omega_{\mu+a}\arr e^{-\s} \left( \omega_{\mu+a} +\le - v^\nu  e^a_\mu 
+e^\nu_a A_\mu \ri  \p_\nu \s \right)  \, ,
\nl
\omega_{\mu a b} & \arr &
\omega_{\mu a b}+  \le 
 e^a_\mu e^\nu_b 
-e^b_\mu e^\nu_a 
\ri \p_\nu \s \,  .
\eea

 In the usual relativistic case the components 
 of a Dirac spinor have all the same Weyl weight;
 this is not true in the null reduction setting that we are considering,
 because the tetrads have different Weyl weights.
 The transformation of the $(\chi,\varphi)$ components is as follows:
 \beq
\chi \arr e^{-2 \s} \chi \, , \qquad \varphi \arr e^{-\s} \varphi \, .
\label{weyl-weight-fermions}
\eeq
 This can be derived from dimensional analysis in the flat case,
see eq.~(\ref{eq-free}): in units of length, $[\varphi] = -1$  and $[\chi] = -2 $.
In the case of a Dirac fermion
\beq
\Psi = \begin{pmatrix}  \chi_L \\ \varphi_L \\ \varphi_R \\ \chi_R    \end{pmatrix} \, ,
\qquad  {\rm dimensions \,\,\, are } \qquad
[\Psi ] = \begin{pmatrix} -2 \\ -1 \\ -1 \\ -2   \end{pmatrix} \, .
\eeq
We note that this Weyl weight choice
 is crucial in order to assign to the term $\bar{\Psi} \Psi $
a well-defined Weyl weight. 
A conformal coupling term such as
$R \bar{\Psi} \Psi$ would have mass dimension $5$,
spoiling conformal invariance.

Promoting $\s$ to a spacetime-dependent function in eq.~(\ref{weyl-weight-fermions}),
one can then verify the Weyl invariance of the action in eq.~(\ref{Lagrangiana-ferm}) 
by direct calculation, using the  non-homogeneus part of the variation of the spin connection
(see eq.~\ref{weyl-spinc}).
One can also check that this is consistent with eq.~(\ref{chichi}):
if we insert $\varphi \arr e^{-\s} \varphi$, we indeed find that
$\chi \arr e^{-2 \s} \chi$.


\section{Heat kernel for fermions}
\label{sect-heat}

\subsection{General framework}

For a complex field $\phi$, the  vacuum functional $W$ is defined by 
\beq
e^{i W} = \int {\cal D}\phi^\dagger\,  {\cal D}\phi\  e^{i S_D[\phi^\dagger,\,\phi]}
\eeq
where $S_D$ is the classical action specified by a differential operator $D$.
In the bosonic case, the path integral is evaluated in terms of the functional determinant of the operator $D$ as
\beq
\label{eqdet}
i W= -\log \det (D)
\eeq
In the fermionic case there are two differencies: 
first the change of sign in the r.h.s. of (\ref{eqdet}) due to the  Berezin functional integration.
 Second, there is the difficulty that the Dirac operator $\slashed D$ is not elliptic after a Wick rotation. 
 This problem  can be bypassed by evaluating  the  determinant 
 of the square of the Dirac operator  and inserting a factor $1/2$:
\beq
\label{eqdet2}
i W= \frac12 \log \det (\slashed D^2)
\eeq
In this way the Euclidean version of the 
squared Dirac operator is elliptic and meets the requirements needed in order to make the heat kernel computation. In fact, 
using anticommutation rules for the product of totally antisymmetric Dirac matrices we find (see e.g. \cite{Christensen:1978md},  \cite{Freedman}):
\beq
 \left( i \slashed D \right)^2  =  -\square + \frac{1}{4} R \equiv - \hat{ \triangle}\, , \qquad
 \Box=D_A D^A \, . 
 \label{dirac2}
\eeq
We need to compute the HK with the Euclidean version of the
operator $ \hat{ \triangle}$ in eq. (\ref{dirac2}). To this purpose we decompose it as
the flat part $\triangle$ plus curved space perturbation $\delta \triangle$:
\beq
\hat{ \triangle}=\triangle \,{  \bf{1}} +\delta \triangle \, ,
\qquad
 \bigtriangleup = \left( - 2 i m \p_t + \p_i^2 \right) \, .
\label{triangolo}
\eeq

\subsection{The flat case}
The computation of the HK is performed 
in Euclidean space. 
This is realized by the substitutions
\beq
t \rightarrow -i t_E \, , \qquad \p_t \rightarrow i \p_{t_E} \, , \qquad m \rightarrow im_E \, .
\label{wick-rotation}
\eeq 
The HK operator of a general euclidean operator ${\hat {\cal O}}_E$  is defined as 
\beq
\hat{K}_{{\hat{{\cal O}}_E}}(s)=\exp (s {\hat{{\cal O}}_E} ) \, .
\eeq
We will denote by $K_{{\hat{{\cal O}}_E}}$ the matrix elements
\beq
K_{{\hat{{\cal O}}_E}}(s,x,t,x',t') = \langle x  t | \hat{K}_{{\hat{{\cal O}}_E}}(s)| x' t' \rangle \, ,
\eeq
and by by $\tilde{K}_{{\hat{{\cal O}}_E}}$ the diagonal matrix elements
\beq
\tilde{K}_{{\hat{{\cal O}}_E}}(s,x,t) = \langle x  t | \hat{K}_{{\hat{{\cal O}}_E}}(s)| x t \rangle \, .
\eeq

In the flat non-relativistic case,
with operator $\triangle$ in (\ref{triangolo}):
\beq
 \bigtriangleup = \left( - 2 i m \p_t + \p_i^2 \right) =
\left( - 2  m \sqrt{- \p_t^2} + \p_i^2 \right) \, ,
\eeq
the heat kernel has been evaluated in \cite{Auzzi:2016lxb} and its matrix elements read
\beq
\label{heatfree}
K_\triangle(s)=\langle x  t | e^{s \triangle} | x' t' \rangle=
\frac{1}{2 \pi} \, \frac{ms}{m^2 s^2 +\frac{(t-t')^2}{4}}  \,
\frac{1}{(4 \pi s)^{d/2} } \exp \le -\frac{(x-x')^2}{4 s} \ri \, .
\eeq
Here a comment is in order: to use the heat kernel machinery with the  
Schr\"odinger operator, 
we  use the formal 
replacement $- 2 i m \p_t \longrightarrow - 2 m \sqrt{- \p_t^2}$. This, by itself, does not render the Schr\"odinger
operator elliptic, but it makes  possible an integral representation in which the exponential of the Schr\"odinger operator 
is written as a sum of exponentials of elliptic operators, which is precisely what is needed to compute the heat kernel, namely
\beq
\label{tria2}
e^{-2 m   \sqrt{-\p_t^2}  }
=\int_0^\infty d \s \frac{m}{\sqrt{\pi}} 
\frac{1}{\s^{3/2}} e^{-\frac{ m^2}{\s}} e^{-\s (-\p_t^2)} \, ,
\eeq
This trick was first introduced  in \cite{Solodukhin:2009sk}, although  in a different context, and used in
 \cite{Auzzi:2016lxb} to 
evaluate the anomaly in the bosonic case. In its essence, this regularization is not different from the one
 normally  used in the relativistic case to adapt the heat kernel procedure to fermions: to make elliptic 
 the Dirac operator, one first considers its square, perform the heat kernel, and then takes the square root
 of the resulting determinant.

\subsection{The curved case}

In order to explicitly compute the functional determinant
we work in coordinate representation with scalar product:
\beq
\langle xt | x' t' \rangle_g = \frac{\delta(x-x')\delta(t-t')}{\sqrt{g}} \, . 
\eeq
In curved background, the HK can be evaluated as a perturbative expansion around (\ref{heatfree}). 
The diagonal matrix elements in the coordinate basis  of the heat kernel
 can be expanded 
in powers of $s$ as:
\beq
\tilde{K}_{\hat{ \triangle}}(s)=
\langle x  t | e^{s \hat{ \triangle}} | x t \rangle_g=
\frac{1}{s^{d/2+1}} 
\le a_0(\hat{ \triangle}) +a_2(\hat{ \triangle}) s + a_4(\hat{ \triangle}) 
s^2 + \dots \ri \, .
\eeq
This defines the De~Witt-Seeley-Gilkey coefficients $a_{2k}(\hat{ \triangle})$ of the problem.
In non-relativistic $2+1$ dimensional theories, the trace anomaly 
is proportional to the $a_4$ coefficient \cite{Auzzi:2016lxb}.

It is convenient to introduce a quantum mechanical space, with flat inner product  
\beq
\langle xt | x' t' \rangle = \delta(x-x')\delta(t-t') \, ,
\eeq
and, for any operator $\hat{\cal O}$, to define the operator $ \hat{M}_{\hat{\cal O}} $ such that
\beq
\langle xt | \hat{\cal O} | x't' \rangle_g = \langle xt | \hat{M}_{\hat{\cal O}} |x' t' \rangle \, .
\eeq
Thus, one introduces and  ``effective'' operator $ \hat{M}_{\hat{\cal O}}$
that keeps track of the metric in the inner product.
In our case, if $\hat{\cal O} =\hat{ \triangle}=  \square - \frac{1}{4} R$, then
\beq
\langle xt | \hat{M} | x't' \rangle = g^{1/4} (x, t) \left( \square_{x,t} - \frac{1}{4} R \right) \left[ g^{-1/4}(x,t) \delta (x-x') \delta (t-t')\right] \, .
\eeq
In this way we can  expand the diagonal elements of 
the HK as
\beq
\tilde{K}_{\hat{M}} (s) = \langle xt | e ^{s \hat{M}} | x t \rangle =
\frac{1}{s^{d/2+1}} \left[ a_0 ( \hat{M}) + s \, a_2 (\hat{M}) + s^2 a_4 (\hat{M}) + \dots \right] \equiv
\sqrt{g} \tilde{K}_{\hat{\bigtriangleup}} (s)  \, .
\eeq

\subsection{A specific perturbation of flat spacetime}
To proceed, we specialize to a particular perturbation of  flat spacetime:
\beq
n_{\mu} = \left( \frac{1}{1- \eta(x^{i})} , 0, 0 \right) \, , \qquad 
v^{\mu} = \left( 1- \eta(x^{i}) , 0, 0 \right) \, , \qquad 
h_{ij}= \delta_{ij}  \, , \qquad   A_{\mu} = 0 \, .
\eeq
Also we remind that 
the spatial frame field is a Kronecker delta:
\beq
e^a_i=e^i_a=\delta^i_a \, .
\eeq
For simplicity, we choose $\eta$ independent from the time coordinate. 
The non-vanishing components of the spin connection and Cristoffel symbols are:
\bea
\omega_{\underset{(M)}{-} + a}  = \frac{1}{2} \frac{\p_a \eta}{1-\eta}   \, , \qquad 
\omega_{\mu \underset{(A)}{-}\underset{(A)}{+} } = - \frac{1}{2} \delta_{\mu i} \frac{\p_i \eta}{1- \eta}  \, , \quad
\omega_{\mu \underset{(A)}{-} a } = \frac{1}{2} \delta_{\mu +} \frac{\p_a \eta}{(1- \eta)^2}  \, ,
\nl
\Gamma^{-}_{\,\,\, \mu -} = \frac{1}{2} \delta_{\mu i} \frac{\p_i \eta}{1- \eta}   \, , \quad
\Gamma^{\rho}_{\,\,\, \mu-} =  - \frac{1}{2}  \delta_{\mu +} \delta^{\rho i} \frac{\p_i \eta}{(1- \eta)^2}  \, , \quad
\Gamma^{\rho}_{\,\,\, \mu \nu} = \frac{1}{2} \delta^{\rho +} \delta_{\mu +} \delta_{\nu i} \frac{\p_i \eta}{1- \eta}  \, .
\eea
The euclidean version of operator $ \hat{M}_{\hat{\cal O}}$ is
obtained by using eq.~(\ref{wick-rotation}): 
\bea
& & g^{1/4} \left( \square - \frac{1}{4} R \right)_{\! \! \!E} g^{-1/4} \Psi    =  \nl   
& =&   \left[ -2 i m  \, \bold{1} +2 i m \eta   \, \bold{1} + \frac{i}{2} (\p_a \eta) \gamma^{+a} \right] \p_t \Psi
  + \left[ -\frac{1}{2} (\p_a \eta) \gamma^{-+}  -\frac{1}{2} \eta (\p_a \eta) \gamma^{-+} \right] \p_a \Psi + \nl
&+ &  \left[ \frac{1}{8} (\p_a \eta)^2 \, \bold{1}- \frac{1}{4} \p^2 \eta \gamma^{-+} - \frac{1}{4} \eta (\p^2 \eta) \gamma^{-+} 
- \frac{1}{4} (\p_a \eta)^2 \gamma^{-+} 
 - \frac{1}{2} m (\p_a \eta) \gamma^{-a} - \frac{1}{2} m \eta (\p_a \eta) \gamma^{-a} \right] \Psi 
+ \nl
& +&  \left[ \frac{1}{16} (\p_a \eta)^2  \bold{1}
+ \frac{1}{16} (\p_a \eta)(\p_b \eta) \lbrace \gamma^{-a} , \gamma^{+b} \rbrace \right] \Psi + \p^2 \Psi \, .
\eea

We will need the matrix elements of $\hat{M}$ in coordinate representation;
to this purpose, it is useful to use the following decomposition:
\beq
\begin{aligned}
\langle x t | \hat{M} | x' t' \rangle  = &
\langle x t | \left[ \bigtriangleup \, \bold{1} + P(x)  \delta(x-x') \delta(t-t')  + S(x) \sqrt{- \p_t^2} \delta(x-x') \delta(t-t') +  \right. \\
& \left. + a_i (x) \, \p_i  \delta(x-x') \delta(t-t')    \right]  | x' t' \rangle  \, ,
\end{aligned} 
\label{Operatore heat kernel}
\eeq
where
\beq
\begin{aligned}
P(x) & =   \frac{3}{16} (\p_i \eta)^2 \, \bold{1} - \frac{1}{4} (\p^2 \eta) \gamma^{-+} - \frac{1}{4} \eta (\p^2 \eta) \gamma^{-+} 
- \frac{1}{4} (\p_i \eta)^2 \gamma^{-+}  \\
& - \frac{1}{2} m (\p_i \eta) \gamma^{-i} - \frac{1}{2} m \eta (\p_i \eta) \gamma^{-i} 
+ \frac{1}{16} (\p_i \eta)(\p_j \eta) \lbrace \gamma^{-i} , \gamma^{-j} \rbrace \, ,  \\
S(x) & = 2 m \eta \, \bold{1} + \frac{1}{2} (\p_i \eta) \gamma^{+i} \, ,  \\
a_i (x) & = - \frac{1}{2} (\p_i \eta) \gamma^{-+} - \frac{1}{2} \eta (\p_i \eta) \gamma^{-+} \, .
\end{aligned} 
\label{Parti heat kernel}
\eeq

A more explicit form is:
\beq
P(x)= \begin{pmatrix}
P_{11} (x) & 0 & 0 & 0 \\ P_{21} (x) & P_{22} (x) & 0 & 0 \\ 0 & 0 & P_{22} (x) & P_{32}(x) \\ 0 & 0 & 0 & P_{11} (x) 
\end{pmatrix} \, ,
\eeq
where
\beq
\begin{aligned}
 P_{11} (x) & = \frac{5}{16} (\p_i \eta)^2 + \frac{1}{4} (\p^2 \eta) + \frac{1}{4} \eta (\p^2 \eta) \, , \\
  P_{22} (x)  & = -\frac{3}{16} (\p_i \eta)^2 - \frac{1}{4} (\p^2 \eta) - \frac{1}{4} \eta (\p^2 \eta) \, ,  \\
P_{21} (x) & =  \frac{\sqrt{2}}{2} m \left[ (\p_1 + i \p_2) \eta + \eta  (\p_1 + i \p_2) \eta \right] \, ,  \\
P_{32} (x) & = \frac{\sqrt{2}}{2} m \left[ (-\p_1 + i \p_2) \eta + \eta  (-\p_1 + i \p_2) \eta \right]  \, .
\end{aligned}
\eeq
Moreover:
\beq
S(x)= \begin{pmatrix}
S_{11} (x) & S_{12} (x) & 0 & 0 \\ 0 & S_{11} (x) & 0 & 0 \\ 0 & 0 & S_{11} (x) & 0 \\ 0 & 0 & S_{43} (x) & S_{11} (x) 
\end{pmatrix} \, ,
\eeq
where
\beq
S_{11} (x) = 2 m \eta \, , \qquad
S_{12} (x) = \frac{\sqrt{2}}{2} (\p_1 - i \p_2 ) \eta \, , \qquad
S_{43} (x) = -  \frac{\sqrt{2}}{2} (\p_1 + i \p_2 ) \eta \, ,
\eeq
and:
\beq
a_i (x)= a_{11} (x) \begin{pmatrix}
1 & 0 & 0 & 0 \\ 0 & -1 & 0 & 0 \\ 0 & 0 & -1 & 0 \\ 0 & 0 & 0 & 1
\end{pmatrix} \, ,
\qquad
a_{11} (x)= \frac{1}{2} (\p_i \eta) +  \frac{1}{2} \eta (\p_i \eta) \, .
\eeq

\subsection{Perturbative expansion}

The next task is to obtain the De~Witt-Seeley-Gilkey
 expansion of the HK operator, in order to find the $ a_4 $ coefficient and then the trace anomaly.
We will split $\hat{M}$  in a free part
plus a perturbation $\hat{V}$:
\beq
 \langle xt | \hat{M} | x' t' \rangle =  \langle xt | \bigtriangleup \, \bold{1} + \hat{V} | x' t' \rangle  =
g^{1/4} \left( \bigtriangleup \, \bold{1} + \delta \bigtriangleup \right) \left[ g^{-1/4} \delta (x-x') \delta (t-t') \right] \, .
\eeq
We can expand perturbatively the HK as a Dyson series:
\beq
\hat{K}_{\hat{M}} (s) = \exp\left[ s \left( \bigtriangleup \, \bold{1} + \hat{V} \right) \right] = \sum_{n=0}^{\infty} \hat{K}_n (s) \, ,
\eeq
where the terms of the sum are 
\beq
\hat{K}_n (s) = \int_{0}^{s} ds_n \, \int_0^{s_n} ds_{n-1} \dots \int_0^{s_2} ds_1 \, 
e^{(s-s_n)\bigtriangleup \, \bold{1}} \hat{V} e^{(s_n-s_{n-1})\bigtriangleup \, \bold{1}} \hat{V} \dots e^{(s_2-s_1)\bigtriangleup \, \bold{1}} \hat{V} e^{s_1 \bigtriangleup \, \bold{1}} \, .
\eeq
Since we are perturbing around  flat space,  from \cite{Auzzi:2016lxb} we have
\beq
K_{\bigtriangleup} (s) = \langle x t | e^{s \bigtriangleup \, \bold{1}} | x' t' \rangle =
\frac{1}{2 \pi} \frac{ms}{m^2 s^2 + \frac{1}{4} (t-t')^2} \frac{1}{(4 \pi s)^{d/2}} \exp\left[- \frac{(x-x')^2}{4s} \right] \, \bold{1} \, ,
\eeq
which gives
\beq
\tilde{K}_{\bigtriangleup} (s) = \mathrm{Tr} \langle x t | e^{s \bigtriangleup \, \bold{1}} | x t \rangle =
\frac{2}{m (4 \pi s)^{d/2+1}} \, \mathrm{Tr} (\bold{1}) = \frac{8}{m (4 \pi s)^{d/2+1}}   \, .
\eeq

\subsubsection{Single insertion}
At the first order the Dyson series is
\beq
K_{1} (s) = \int_0^s ds' \, \langle xt | e^{(s-s') \bigtriangleup} \hat{V} e^{s' \bigtriangleup} | x' t' \rangle \, .
\eeq
According to eq.~(\ref{Operatore heat kernel}), we can decompose the expression as
\beq
\begin{aligned}
K_{1} (s)  &= K_{1P} (s) + K_{1S} (s) + K_{1a_i} (s)   = \mathrm{Tr} \int_0^s ds' \, \langle xt | e^{(s-s') \bigtriangleup} P(x) e^{s' \bigtriangleup} | x' t' \rangle + \\
& + \mathrm{Tr} \int_0^s ds' \, \langle xt | e^{(s-s') \bigtriangleup} S(x) \sqrt{-\p_t^2} e^{s' \bigtriangleup} | x' t' \rangle  + \mathrm{Tr} \int_0^s ds' \, \langle xt | e^{(s-s') \bigtriangleup} a_i(x) \p_i e^{s' \bigtriangleup} | x' t' \rangle \, .
\end{aligned}
\eeq
The contribution $ K_{1a_i} (s)$ contains an implicit sum over the index $i$.

Note that we also introduced in the expression the trace operation, since we are now dealing with squared matrices and the heat kernel expansion required a trace over the operator considered.

We can use the following results from Appendix A of \cite{Auzzi:2016lxb}:
\beq
\tilde{K}_{1P} = \frac{2}{m(4 \pi s)^{d/2+1}} \mathrm{Tr} \left( s P + \frac{1}{6} s^2 \p_x^2 P + \dots \right) \, ,
\eeq
\beq
\tilde{K}_{1S} = \frac{2}{m(4 \pi s)^{d/2+1}} \mathrm{Tr} \left( \frac{S}{2m} + \frac{s}{12m} \p^2 S + \frac{s^2}{120m} \p^4 S + \dots \right) \, .
\eeq
Moreover the contribution $\tilde{K}_{1a_i}$ due to $a_i$
is the sum of the trace of various terms  proportional 
to the derivatives of $a_i$; these terms have all zero trace and so
$\tilde{K}_{1a_i}=0$.


\subsubsection{Double insertion}

At the second order the heat kernel expansion is
\beq
K_{2} (s) = 
\int_0^s ds' \int_0^{s'} ds''  \, \langle xt | e^{(s-s') \bigtriangleup} \hat{V} e^{(s'-s'') \bigtriangleup}
\hat{V} e^{s'' \bigtriangleup} | x' t' \rangle \, .
\eeq
$K_{2}$ splits into the sum of several contributions:
\beq
K_{2} (s)  = 
 \sum_{X} K_{2X} (s) =
 \eeq
 \[=
K_{2 P P} (s) + K_{2 S S} (s)  +  K_{2 P S} (s) + K_{2 S P} (s) + K_{2a_i a_j} (s) + 
K_{2a_i P} (s) + K_{2P a_i} (s) + K_{2a_i S} (s) + K_{2 S a_i} (s)  \, ,
\]
where in each contribution there is an implicit sum over
the indices $i,j$.

We can use the following results from Appendix B of \cite{Auzzi:2016lxb}:
\beq
\tilde{K}_{2PP} = \frac{2}{m(4 \pi s)^{d/2+1}} \mathrm{Tr} \left( \frac{s^2}{2} P(x)^2 + \dots \right)  \, ,
\eeq
\beq
\begin{aligned}
\tilde{K}_{2SS}  = & \frac{2}{m(4 \pi s)^{d/2+1}} \mathrm{Tr} \left( \frac{S^2}{4m^2} + \frac{s}{12m^2} S \p^2 S +
\frac{s}{24m^2} \p_k S \p_k S + \frac{s^2}{120m^2} S \p^4 S + \right. \\
& \left. +\frac{s^2}{144m^2} \p^2 S \p^2 S +
\frac{s^2}{60m^2} \p_i\p^2 S \p_i S + \frac{s^2}{180m^2} \p_{ij} S \p_{ij} S  + \dots \right)  \, ,
\end{aligned}
\eeq
\beq
\tilde{K}_{2PS} = \tilde{K}_{2SP}   =  \frac{1}{m(4 \pi s)^{d/2+1}} \mathrm{Tr} \left( \frac{s}{2m} SP + \frac{s^2}{12m} S \p^2 P + \frac{s^2}{12m}\p^2 S  P + \frac{s^2}{12m} \p_i S \p_i P + \dots \right)  \, .
\eeq
For the remaining terms, the calculation is performed in Appendix \ref{appe-2insertions}:
\bea
\label{a-p1}
\tilde{K}_{2 a_j a_i } &=& \frac{2 }{m (4 \pi s)^{d/2+1}}  
\textrm{Tr}  \left[ -  \frac{s}{4}  a_i a_i    
- \frac{s^2}{24} (\p_{j} a_i) (\p_i a_j)  \right.
\nl
 & & \left. 
+  \frac{s^2}{8} (\p_{i} a_i) (\p_{j} a_j)   -  \frac{s^2}{12} a_i (\p^2 a_i) 
 -  \frac{s^2}{24} (\p_i a_j)^2 
+  \dots \right] \, ,
\eea
\beq
\label{a-p2}
\tilde{K}_{2 a_i P}  =  \frac{2}{m(4 \pi s)^{d/2+1}} \mathrm{Tr} \left( - \frac{s^2}{3} P (\p_i a_i)  - \frac{s^2}{6} (\p_i P) a_i  + \dots \right)  \, ,
\eeq
\beq
\label{a-p3}
\tilde{K}_{2  P a_i}  =  \frac{2}{m(4 \pi s)^{d/2+1}} \mathrm{Tr} \left(   \frac{s^2}{6} a_i (\p_i P) - \frac{s^2}{6} (\p_i a_i) P + \dots \right)  \, .
\eeq
The expressions for $ \tilde{K}_{2 a_i S}  , \tilde{K}_{2  S a_i}    $  involve 
traces of matrix products of the kind 
\[ {\rm Tr} \,\,\,
\p^k a_i(x) \p^l S(x) \, ,\] 
but these all vanish due to the structure of the matrices $a_i$, $S$,
 whose entries sit in orthogonal subspaces.

\subsubsection{Results}

Summing the contribution from the single and double insertions,
we find $a_4$ up to the second order in $\eta$, for $d=2$:
\beq
 \sqrt{g} a_4(\hat{\triangle}) = 
 \frac{2}{m(4 \pi)^{2}} \left[ \frac{1}{15} \p^4 \eta +  \frac{2}{15} \eta (\p^4 \eta)       
+  \frac{13}{30} (\p_i \eta) (\p_i \p^2 \eta) + \frac{1}{9} (\p^2 \eta)^2 +  \frac{31}{180} (\p_{ij} \eta  )^2  \right] \, . 
\eeq
We should then express the result in terms of curvature invariants.
Up to the second order in $\eta$, the curvature combinations entering  the anomaly are given by:
\bea
\sqrt{g} D^2 R &=& -2 \p^4 \eta -4 \eta (\p^4 \eta)       
-13 (\p_i \eta) (\p_i \p^2 \eta) -  2 (\p^2 \eta)^2 - 7 (\p_{ij} \eta  )^2   \, ,
\nl
\sqrt{g} E_4 &=& 2 (\p^2 \eta)^2 - 2 (\p_{ij} \eta)^2  \, ,
\qquad
\sqrt{g} W^2 = \frac{1}{3} (\p^2 \eta)^2  \, .
\eea
In our conventions the
 Euler density $E_4$ and
 the square of the Weyl tensor $W^2$ are, 
 in term of the Riemann and Ricci tensor
 of the null reduction metric eq.~(\ref{null-red}):
\beq
E_4=R^2_{ABMN} -4 R^2_{AB} +R^2 \, ,
\qquad
W^2_{ABMN}=R^2_{ABMN} -2 R^2_{AB} +\frac{1}{3} R^2 \, .
\eeq

Since we are studying a Weyl-invariant operator, we know from the
Wess-Zumino consistency conditions that the $ R^2 $ term cannot enter the anomaly.
We can then write the result as:
\beq
  a_4 ( \slashed{D}^2_E) = \frac{1}{8m\pi^2} \left( \frac{11}{360} E_4 - \frac{1}{20} W^2  - \frac{1}{30} D^2 R   \right)   \, . 
\eeq
The trace anomaly then can be computed as follows:
\beq
 {\mathcal{A} } = - \frac{1}{2} a_4 (\slashed D_E^2)  \, . 
\eeq

\section{Conclusions}
\label{sect-concl}

In this paper we checked that the action of 
a non-relativistic spin $1/2$ fermion coupled to NC geometry
is Weyl invariant. 
Then the trace anomaly was computed using the HK method; 
the result is for a fermionic spin doublet:
\beq
 \mathcal{A} = T^i_{\,\,\, i} - 2 T^0_{\,\,\, 0} = 
\left( -a \, E_4 + c \, W^2  +b R^2+ a' \, D^2 R  \right) + \dots \,\, . 
\eeq
where
\beq
a=\frac{1}{8 m \pi^2} \frac{1}{360} \, \frac{11}{2} \, , \quad
c=\frac{1}{8 m \pi^2} \frac{9}{360} \, , \quad 
b= 0 \, , \qquad
a'=\frac{1}{8 m \pi^2}  \frac{6}{360} \, ,
\eeq
and the dots stand for possible additional terms,
both higher derivatives and of the
kind discussed in  \cite{Fernandes:2017nvx},
which violate the Milne boost symmetry. 

Up to an overall $1/m$ multiplicative factor,
 the anomaly coefficients turn out to be proportional to
the ones of a relativistic Dirac fermion  in $4$ dimensions.
A similar numerical coincidence happens also in the scalar case
\cite{Auzzi:2016lxb}, where the value of the anomaly coefficients is:
\beq
a=\frac{1}{8 m \pi^2} \frac{1}{360}  \, , \quad
c=\frac{1}{8 m \pi^2} \frac{3}{360} \, , \quad
b=\frac{1}{8 m \pi^2} \frac{1}{2} \le \xi -\frac{1}{6}\ri^2 \, , \quad
a'=\frac{1}{8 m \pi^2}  \frac{1-5 \xi}{30} \, ,
\eeq
where $\xi$ is the parameter multiplying the conformal coupling. 

It is natural to conjecture that an analog of the $a$-theorem
may hold for the $E_4$ coefficient of Schr\"odinger-invariant
theories in $2+1$ dimensions.
For example,
in the case where both the elementary and the composite degrees of freedom
would be free
scalars and fermions with spin $1/2$, it would imply that
\beq
a_{\rm UV} \propto \sum_{\rm scalars}^{\rm UV} \frac{1}{m} +
\frac{11}{2}  \sum_{\rm fermions}^{\rm UV} \frac{1}{m} 
\geq
\sum_{\rm scalars}^{\rm IR} \frac{1}{m} +
\frac{11}{2}  \sum_{\rm fermions}^{\rm IR} \frac{1}{m}
\propto a_{\rm IR} \, .
\label{aaath}
\eeq
 In Galilean-invariant theories the mass
is a conserved quantity and the mass of a bound state is equal to the sum
of the masses of the elementary constituents: no bound-state
contribution to the mass is present as in the relativistic case.
As proposed in \cite{Auzzi:2016lxb}, 
the $1/m$ dependence is consistent with the 
intuition that bound states form in the infrared:
as energy is added bound states tend to be broken. \\
Several interesting problems require further investigation:
\begin{itemize}
\item Some new anomaly terms were computed in \cite{Fernandes:2017nvx}
using Fujikawa approach;
they are present when a non-trivial background $U(1)$ gauge field is added
and they violate Milne boost symmetry.
 Wess-Zumino consistency conditions
 for these new terms should be studied
 and the computation should be checked using HK method.
\item The relation between the anomaly coefficients and the 
correlation functions of the energy-momentum tensor multiplet should be
clarified. In the case of vacuum correlation function, these  correlators 
 have support just at coincident points.
It would be interesting to check if the anomaly coefficients can be related
to the form of the finite-density correlators evaluated at separated points. 
\item The relation between the anomaly and the dilaton effective action
should be investigated; in the relativistic case, 
this leads to a proof of the $a$-theorem \cite{Komargodski:2011vj}. 
The study of non-relativistic dilaton was initiated in \cite{Arav:2017plg}.
\item It would be interesting to attempt a perturbative proof
using Osborn's local renormalization group  approach; 
this was initiated in  \cite{Auzzi:2016lrq}. The main 
missing ingredient to the proof is to control the positivity of
some anomaly coefficients whose relativistic analog turn out
to be proportional to the Zamolodchikov metric.
\item In the relativistic supersymmetric case, 
 there is a powerful relation between the
 trace anomaly coefficients and $R$-charges
  \cite{Anselmi:1997am}; it would be interesting to check if 
  a similar relation exists also in the non-relativistic case.
  The supersymmetric local RG approach
as in \cite{Auzzi:2015yia} might be a convenient way to investigate these issues. 
Newton-Cartan supergravity was studied in \cite{Andringa:2013mma}.
\item The anomaly coefficients for anyons
coupled to NC backgrounds should be computed. This may be interesting for 
condensed matter applications, as the quantum Hall effect.
\end{itemize}


\section*{Appendix}
\addtocontents{toc}{\protect\setcounter{tocdepth}{1}}
\appendix

\section{Sigma and gamma matrices with light-cone indices}
\label{app-gamma}

We use the standard conventions:
\beq
\sigma^{A} = (\bold{1}, \sigma^\a) \, , \qquad  \bar{\sigma}^{A} = (- \bold{1},  \sigma^\a)
\eeq
It is useful to write the explicit expressions in lightcone indices:
\begin{itemize}

\item{Sigma matrices in 4 dimensions 
\bea
\sigma^{\pm} = \frac{1}{\sqrt{2}} (\sigma^{3} \pm \sigma^{0}) \, , &&\qquad
\bar{\sigma}^{\pm} = \frac{1}{\sqrt{2}} (\bar{\sigma}^{3} \pm \bar{\sigma}^{0})  \, ,
\nl
\sigma^- = \sqrt{2}
\begin{pmatrix}
0 & 0 \\
0 & -1 
\end{pmatrix} \, , &&\qquad
\sigma^+ = \sqrt{2}
\begin{pmatrix}
1 & 0 \\
0 & 0 
\end{pmatrix} \, , 
\nl
\bar{\sigma}^- = \sqrt{2}
\begin{pmatrix}
1 & 0 \\
0 & 0 
\end{pmatrix} \, , && \qquad
\bar{\sigma}^+ = \sqrt{2}
\begin{pmatrix}
0 & 0 \\
0 & -1 
\end{pmatrix} \, ,
\nl
\sigma^1 = \bar{\sigma}^1=
\begin{pmatrix}
0 & 1 \\
1 & 0 
\end{pmatrix}  \, , && \qquad
\sigma^2 = \bar{\sigma}^2 =
\begin{pmatrix}
0 & -i \\
i & 0 
\end{pmatrix}  \, ,
\eea
The Lorentz generators are then:
\beq
\sigma^{AB} = \frac{1}{2} \left(  \sigma^A \bar{\sigma}^B  - \sigma^B \bar{\sigma}^A \right) \, ,
\eeq
which gives:
\bea
\sigma^{-1}= - \frac{1}{\sqrt{2}} (\sigma^1 -i \sigma^2) = \sqrt{2} \begin{pmatrix} 0 & 0 \\ -1 & 0    \end{pmatrix} \, , &&\qquad
\sigma^{-2}= - \frac{1}{\sqrt{2}} (\sigma^2 +i \sigma^1) = \sqrt{2} i \begin{pmatrix} 0 & 0 \\ -1 & 0    \end{pmatrix} \, ,
\nl
\sigma^{+1}=  \frac{1}{\sqrt{2}} (\sigma^1 + i \sigma^2) = \sqrt{2} \begin{pmatrix} 0 & 1 \\ 0 & 0    \end{pmatrix} \, , &&\qquad
\sigma^{+2}= - \frac{1}{\sqrt{2}} (i \sigma^1 - \sigma^2) = \sqrt{2} i \begin{pmatrix} 0 & -1 \\ 0 & 0    \end{pmatrix} \, ,
\nl
\sigma^{-+}= -\sigma^3= \begin{pmatrix} -1 & 0 \\ 0 & 1    \end{pmatrix} \, , &&\qquad
\sigma^{12}=  i \sigma^3 = \begin{pmatrix} i & 0 \\ 0 & - i    \end{pmatrix} \, .
\eea }

\item{ Gamma matrices in 4 dimensions 
\bea
\gamma^{-}=\frac{1}{\sqrt{2}} (\gamma^{3}- \gamma^0) = \sqrt{2}  \begin{pmatrix}
0 & 0 &  0 & 0 \\
0 & 0 &  0 & -1 \\
1 & 0 &  0 & 0 \\
0 & 0 &  0 & 0 \\  
\end{pmatrix} \, , &&
 \quad \gamma^{+}= \frac{1}{\sqrt{2}} (\gamma^{3}+ \gamma^0) = \sqrt{2}  \begin{pmatrix}
0 & 0 &  1 & 0 \\
0 & 0 &  0 & 0 \\
0 & 0 &  0 & 0 \\
0 & -1 &  0 & 0 \\  
\end{pmatrix}  \, ,
\nl
\gamma^1 = \begin{pmatrix}
0 & \sigma^1 \\
 \sigma^1 & 0  
\end{pmatrix}  \, , && \qquad
\gamma^2 = \begin{pmatrix}
0 & \sigma^2 \\
 \sigma^2 & 0  
\end{pmatrix}  \, ,
\eea}
The Lorentz generators are:
\beq
\gamma^{AB}= \frac{1}{2} [\gamma^A , \gamma^B ] \, ,
\eeq
which gives:
\bea
\gamma^{-+} &=&  \begin{pmatrix}
-1 & 0 & 0 & 0 \\ 0 & 1 & 0 & 0 \\ 0 & 0 & 1 & 0 \\ 0 & 0 & 0 & -1
\end{pmatrix} \, , \quad
\gamma^{-1} =  \begin{pmatrix}
0 & 0 & 0 & 0 \\ -\sqrt{2} & 0 & 0 & 0 \\ 0 & 0 & 0 & \sqrt{2} \\ 0 & 0 & 0 & 0
\end{pmatrix} \, , \quad
\gamma^{-2} = \begin{pmatrix}
0 & 0 & 0 & 0 \\ -\sqrt{2} i & 0 & 0 & 0 \\ 0 & 0 & 0 & - \sqrt{2} i \\ 0 & 0 & 0 & 0
\end{pmatrix}  \, , 
\nl
\gamma^{+1} &=&  \begin{pmatrix}
0 & \sqrt{2} & 0 & 0 \\ 0 & 0 & 0 & 0 \\ 0 & 0 & 0 & 0 \\ 0 & 0 & -\sqrt{2} & 0
\end{pmatrix} \, , \quad
\gamma^{+2} = \begin{pmatrix}
0 & -\sqrt{2} i & 0 & 0 \\ 0 & 0 & 0 & 0 \\ 0 & 0 & 0 & 0 \\ 0 & 0 & -\sqrt{2} i & 0
\end{pmatrix}    \, .
\eea
\end{itemize}


\section{Spin connection}
\label{spinc}

The explicit expression for the spin connection is:
\bea
\omega_{MAB} &=& \frac{1}{2} \left[ e^{N}_{\,\,\,A} \left( \p_M  e_{NB} - \p_N   e_{MB} \right) 
-  e^{N}_{\,\,\,B}  \left( \p_M  e_{NA} - \p_N   e_{MA} \right)  \right. \nl
& & \left. 
- e^{N}_{\,\,\,A} e^{P}_{\,\,\,B}  \left( \p_N  e_{PC} - \p_P   e_{NC} \right)   e^{C}_{\,\,\,M} \right] \, .
\eea
We thus obtain the components:
\bea
\omega_{\underset{(M)}{-} AB} &=& - \frac{1}{2} e^{\mu}_{\,\,\,A} e^{\nu}_{\,\,\, B} \tilde{F}_{\mu\nu} \, , \quad
\omega_{\mu \underset{(A)}{-} A}   =  -\frac{1}{2}  e^{\nu}_{\,\,\,A}  \tilde{F}_{\mu\nu}  \, , 
\nl
\omega_{\mu \underset{(A)}{+} a} &=& \frac{1}{2} v^{\nu} \left(  \p_{\mu} e^{a}_{\,\,\, \nu} -  \p_{\nu} e^{a}_{\,\,\, \mu}  \right)  - \frac{1}{2}   e^{\nu}_{\,\,\, a} F_{\mu\nu} - \frac{1}{2} v^{\nu} e^{\rho}_{\,\,\, a}  \left[  A_{\mu}  \tilde{F}_{\nu\rho} + n_{\mu} F_{\nu\rho} +e^b_{\,\,\, \mu} \left( \p_{\nu} e^{b}_{\,\,\, \rho}-  \p_{\rho} e^{b}_{\,\,\, \nu}  \right)  \right] \, , 
\nl
\omega_{\mu  a b}  &=& \frac{1}{2} e^{\nu}_{\,\,\,a} \left(  \p_{\mu} e^{b}_{\,\,\, \nu} -  \p_{\nu} e^{b}_{\,\,\, \mu}  \right)  - \frac{1}{2}   e^{\nu}_{\,\,\, b} \left(  \p_{\mu} e^{a}_{\,\,\, \nu} -  \p_{\nu} e^{a}_{\,\,\, \mu}   \right) + 
\nl
& & - \frac{1}{2}e^{\nu}_{\,\,\,a} e^{\rho}_{\,\,\, b}  \left[  A_{\mu}  \tilde{F}_{\nu\rho} + n_{\mu} F_{\nu\rho} +e^c_{\,\,\, \mu} \left( \p_{\nu} e^{c}_{\,\,\, \rho}-  \p_{\rho} e^{c}_{\,\,\, \nu}  \right)  \right]  \, . 
\eea
Note that $\omega_{\underset{(M)}{-} -B}=0$.


\section{Some double insertion contributions to the Heat Kernel}
\label{appe-2insertions}

Here we consider contributions of the form  $K_{2 X_1 X_2}(s)$, where
\beq
X_1=\left\{ P(x_1), a_i(x_1)  \right\} \, , 
\qquad
X_2=\left\{ P(x_2), a_j(x_2)  \right\} \, . 
\eeq
 whose explicit expression is:
\bea
K_{2 X_1 X_2}(s) &=&
\int_0^s d s_2 \int_0^{s_2} ds_1
\nl
& &
\langle x' t' | e^{-(s-s_2)\triangle}
| x_2 t_2 \rangle 
\hat{X}_2
\langle x_2 t_2 | e^{-(s_2-s_1) \triangle}  | x_1 t_1 \rangle
\hat{X}_1
\langle 
x_1 t_1 | e^{-s_1 \Delta }
| x t \rangle \, ,
\eea
where
\beq
\hat{X}_1=\left\{ P(x_1), a_i(x_1)  \p_{x_1,i} \right\} \, , 
\qquad
\hat{X}_2=\left\{ P(x_2), a_j(x_2) \p_{x_2,j} \right\} \, . 
\eeq
The quantity $K_{2PP}$ was already computed in \cite{Auzzi:2016lxb}.

We can split the integration as follows:
\beq
K_{2 X_1 X_2}(s)=\int_0^s d s_2 \int_0^{s_2} ds_1
\frac{1}{(4 \pi (s-s_2))^{d/2} } \, 
\frac{1}{(4 \pi (s_2-s_1))^{d/2} }  \, 
\frac{1}{(4 \pi s_1)^{d/2} } 
\Xi^{X_1 X_2} \,  
\Theta \, ,
\eeq
where $\Xi^{X_1 X_2}$ and $\Theta$
correspond to the space and time integrals, respectively.
It is useful to Fourier transform:
\beq
\Xi^{X_1 X_2}=
\int \frac{d^d k_1}{(2 \pi)^{d/2}}
\frac{d^d k_2}{(2 \pi)^{d/2}}
\tilde{\Xi}^{X_1 X_2} \, ,
\eeq
and to introduce:
\beq
\Upsilon=
\exp \le  i k_1 x_1+i k_2 x_2 
 -\frac{(x'-x_2)^2}{4 (s-s_2)}
 -\frac{(x_2-x_1)^2}{4 (s_2-s_1)} 
-\frac{(x_1-x)^2}{4 s_1}\ri \, .
\eeq
The Fourier transforms of the space part of the integrals are:
\bea
 \tilde{\Xi}^{PP} &=&
\int d x_1 \int d x_2 \Upsilon P(k_1) P(k_2)  \, ,
\nl 
\tilde{\Xi}^{a_i P} &=&
-\p_{x,i}\left[  \int d x_1 \int d x_2 
\Upsilon a_i(k_1) P(k_2) \right] \, ,
\nl
\tilde{\Xi}^{P a_j } &=&
 \int d x_1 \int d x_2 
 \left[ -\frac{(x_2-x_1)_j}{2 (s_2-s_1)} \right]
\Upsilon P(k_1) a_j(k_2)  \, ,
\nl
\tilde{\Xi}^{a_i a_j }&=&
-\p_{x,i} \left[
 \int d x_1 \int d x_2 
 \left[ -\frac{(x_2-x_1)_j}{2 (s_2-s_1)} \right]
\Upsilon a_i(k_1) a_j(k_2)  \right] \, ,
\eea
where $P(k)$ and $a_i(k)$ are the Fourier transform of
$P(x)$ and $a_i(x)$.
The two basic integrals give:
\begin{eqnarray*}
 \tilde{\Xi}^{PP} &=&
 (4 \pi)^{d} \le \frac{s_1 (s-s_2) (s_2-s_1)}{s} \ri^{d/2}
 \nl
 & &
\exp \le 
\frac{i k_1 s_1 x'}{s}+\frac{i k_2 s_2 x'}{s}-\frac{i k_1 s_1 x}{s}
-\frac{i k_2 s_2 x}{s}+\frac{k_1^2 s_1^2}{s}
+\frac{k_2^2 s_2^2}{s}
-k_1^2   s_1-2 k_1 k_2 s_1-k_2^2 s_2
\right.
\nl & &
\left.
+\frac{2 k_1 k_2 s_1 s_2}{s}+i k_1 x+i k_2 x
-\frac{x^2}{4 s}+\frac{x x'}{2 s}-\frac{\left(x'\right)^2}{4 s}
\ri P(k_1) P(k_2)  \, ,
\end{eqnarray*}
\begin{eqnarray*}
 \tilde{\Xi}^{Pa_j} &=&
 \exp \le
\frac{i k_1 s_1 x'}{s}+\frac{i k_2 s_2 x'}{s}-\frac{i k_1 s_1 x}{s}
-\frac{i k_2 s_2 x}{s}+\frac{k_1^2 s_1^2}{s}+\frac{k_2^2
   s_2^2}{s}-k_1^2 s_1-2 k_1 k_2 s_1-k_2^2 s_2
  \right.
\nl & &
  \left.
   + \frac{2 k_1 k_2 s_1 s_2}{s}+i k_1 x+i k_2 x+\frac{x x'}{2
   s}-\frac{\left(x'\right)^2}{4 s}-\frac{x^2}{4 s} \ri 
    (4 \pi)^{d}  \le \frac{ s_1 \left(s_1-s_2\right) \left(s_2-s\right)}{s} \ri^{d/2}
  \nl  & &
     \frac{ \left( i k_1 s_1+  i k_2 s_2 - i k_2 s
   +\frac{x-x'}{2} \right)_i}{s}   P(k_1)  a_j(k_2) \, .
\end{eqnarray*}
The expressions for 
$\tilde{\Xi}^{a_i P} $ and $\tilde{\Xi}^{a_i a_j }$
can be obtained differentiating
  $\tilde{\Xi}^{PP}$ and $ \tilde{\Xi}^{Pa_j}$ with respect to $x_i$.
The time part gives:
\bea
\Theta &=&
\frac{1}{(2  \pi)^3}
 \int \! d t_1  \int \!  d t_2
 \frac{m (s-s_2)}{m^2 (s-s_2)^2 +\frac{(t_2-t')^2}{4}} \,
 \frac{m (s_2-s_1)}{m^2 (s_2-s_1)^2 +\frac{(t_2-t_1)^2}{4}} \,
 \frac{ms_1}{m^2 s_1^2 +\frac{(t_1-t)^2}{4}} =
\nl
&=& \frac{1}{\pi} \frac{2 m s}{4 m^2 s^2+\left(t-t'\right)^2} \, .
\eea
Combining all the expressions and specializing to $x=x'$, $t=t'$,
we find eqs.~(\ref{a-p1})-(\ref{a-p3}).


\begin{thebibliography}{1}


\bibitem{gravitation}
 Charles W. Misner, Kip S. Thorne and John Archibald Wheeler (1973), Gravitation,
 San Francisco: W. H. Freeman, ISBN 978-0-7167-0344-0.


\bibitem{Son:2005rv}
  D.~T.~Son and M.~Wingate,
  Annals Phys.\  {\bf 321} (2006) 197
  [cond-mat/0509786].

\bibitem{Hoyos:2011ez}
  C.~Hoyos and D.~T.~Son,
  Phys.\ Rev.\ Lett.\  {\bf 108} (2012) 066805
  [arXiv:1109.2651 [cond-mat.mes-hall]].

\bibitem{Son:2013rqa}
  D.~T.~Son,
  arXiv:1306.0638 [cond-mat.mes-hall].

\bibitem{Geracie:2014nka}
  M.~Geracie, D.~T.~Son, C.~Wu and S.~F.~Wu,
  Phys.\ Rev.\ D {\bf 91} (2015) 045030
  [arXiv:1407.1252 [cond-mat.mes-hall]].

  
  \bibitem{Zamolodchikov:1986gt}
    A.~B.~Zamolodchikov,
    JETP Lett.\  {\bf 43} (1986) 730
     [Pisma Zh.\ Eksp.\ Teor.\ Fiz.\  {\bf 43} (1986) 565].
 

 
 \bibitem{Jafferis:2010un}
  D.~L.~Jafferis,
  JHEP {\bf 1205} (2012) 159
  doi:10.1007/JHEP05(2012)159
  [arXiv:1012.3210 [hep-th]].

\bibitem{Jafferis:2011zi}
  D.~L.~Jafferis, I.~R.~Klebanov, S.~S.~Pufu and B.~R.~Safdi,
  JHEP {\bf 1106} (2011) 102
  doi:10.1007/JHEP06(2011)102
  [arXiv:1103.1181 [hep-th]].

\bibitem{Pufu:2016zxm}
  S.~S.~Pufu,
  arXiv:1608.02960 [hep-th].

 
 \bibitem{Cardy:1988cwa}
  J.~L.~Cardy,
  Phys.\ Lett.\ B {\bf 215} (1988) 749.

 \bibitem{Osborn:1989td}
  H.~Osborn,
  Phys.\ Lett.\ B {\bf 222} (1989) 97.
 
 \bibitem{Jack:1990eb}
  I.~Jack and H.~Osborn,
  Nucl.\ Phys.\ B {\bf 343} (1990) 647.
 
 \bibitem{Osborn:1991gm}
  H.~Osborn,
  Nucl.\ Phys.\ B {\bf 363} (1991) 486.
  
  
\bibitem{Komargodski:2011vj}
  Z.~Komargodski and A.~Schwimmer,
  JHEP {\bf 1112} (2011) 099
  [arXiv:1107.3987 [hep-th]].
 
 \bibitem{Komargodski:2011xv}
  Z.~Komargodski,
  JHEP {\bf 1207} (2012) 069
  [arXiv:1112.4538 [hep-th]].
  
  
\bibitem{Deser:1993yx}
  S.~Deser and A.~Schwimmer,
  Phys.\ Lett.\ B {\bf 309} (1993) 279
  [hep-th/9302047].

 
\bibitem{Adam:2009gq}
  I.~Adam, I.~V.~Melnikov and S.~Theisen,
  JHEP {\bf 0909} (2009) 130
  [arXiv:0907.2156 [hep-th]].

\bibitem{Baggio:2011ha}
  M.~Baggio, J.~de Boer and K.~Holsheimer,
  JHEP {\bf 1207} (2012) 099
  [arXiv:1112.6416 [hep-th]].

\bibitem{Griffin:2011xs}
  T.~Griffin, P.~Horava and C.~M.~Melby-Thompson,
  JHEP {\bf 1205} (2012) 010
  [arXiv:1112.5660 [hep-th]].


 
\bibitem{Arav:2014goa}
  I.~Arav, S.~Chapman and Y.~Oz,
  JHEP {\bf 1502} (2015) 078
  doi:10.1007/JHEP02(2015)078
  [arXiv:1410.5831 [hep-th]].

\bibitem{Arav:2016xjc}
  I.~Arav, S.~Chapman and Y.~Oz,
  JHEP {\bf 1606} (2016) 158
  doi:10.1007/JHEP06(2016)158
  [arXiv:1601.06795 [hep-th]].
  
\bibitem{Barvinsky:2017mal}
  A.~O.~Barvinsky, D.~Blas, M.~Herrero-Valea, D.~V.~Nesterov, G.~Pérez-Nadal and C.~F.~Steinwachs,
  arXiv:1703.04747 [hep-th].
  
\bibitem{Arav:2016akx}
  I.~Arav, Y.~Oz and A.~Raviv-Moshe,
  JHEP {\bf 1703} (2017) 088
  doi:10.1007/JHEP03(2017)088
  [arXiv:1612.03500 [hep-th]].
  
  
  
\bibitem{Jensen:2014hqa}
  K.~Jensen,
  arXiv:1412.7750 [hep-th].
   
\bibitem{Auzzi:2015fgg}
  R.~Auzzi, S.~Baiguera and G.~Nardelli,
  JHEP {\bf 1602} (2016) 003
   Erratum: [JHEP {\bf 1602} (2016) 177]
  doi:10.1007/JHEP02(2016)003, 10.1007/JHEP02(2016)177
  [arXiv:1511.08150 [hep-th]].
  
\bibitem{Auzzi:2016lrq}
  R.~Auzzi, S.~Baiguera, F.~Filippini and G.~Nardelli,
  JHEP {\bf 1611} (2016) 163
  doi:10.1007/JHEP11(2016)163
  [arXiv:1610.00123 [hep-th]].
  
\bibitem{Auzzi:2016lxb}
  R.~Auzzi and G.~Nardelli,
  JHEP {\bf 1607} (2016) 047
  doi:10.1007/JHEP07(2016)047
  [arXiv:1605.08684 [hep-th]].

\bibitem{Fernandes:2017nvx} 
  K.~Fernandes and A.~Mitra,
  arXiv:1703.09162 [gr-qc].

\bibitem{Pal:2017ntk} 
  S.~Pal and B.~Grinstein,
  arXiv:1703.02987 [hep-th].


\bibitem{Duval:1984cj}
  C.~Duval, G.~Burdet, H.~P.~Kunzle and M.~Perrin,
  Phys.\ Rev.\ D {\bf 31} (1985) 1841.


\bibitem{Jensen:2014aia}
  K.~Jensen,
  arXiv:1408.6855 [hep-th].
  
\bibitem{Hartong:2014pma}
  J.~Hartong, E.~Kiritsis and N.~A.~Obers,
  Phys.\ Rev.\ D {\bf 92} (2015) 066003
  doi:10.1103/PhysRevD.92.066003
  [arXiv:1409.1522 [hep-th]].
  
\bibitem{Hartong:2014oma}
  J.~Hartong, E.~Kiritsis and N.~A.~Obers,
  Phys.\ Lett.\ B {\bf 746} (2015) 318
  doi:10.1016/j.physletb.2015.05.010
  [arXiv:1409.1519 [hep-th]].
  
\bibitem{Hartong:2015wxa}
  J.~Hartong, E.~Kiritsis and N.~A.~Obers,
  JHEP {\bf 1508} (2015) 006
  doi:10.1007/JHEP08(2015)006
  [arXiv:1502.00228 [hep-th]].

  
\bibitem{Geracie:2015dea}
  M.~Geracie, K.~Prabhu and M.~M.~Roberts,
  J.\ Math.\ Phys.\  {\bf 56} (2015) no.10,  103505
  doi:10.1063/1.4932967
  [arXiv:1503.02682 [hep-th]].

\bibitem{Geracie:2015xfa}
  M.~Geracie, K.~Prabhu and M.~M.~Roberts,
  JHEP {\bf 1508} (2015) 042
  doi:10.1007/JHEP08(2015)042
  [arXiv:1503.02680 [hep-th]].



\bibitem{Geracie:2016dpu}
  M.~Geracie, K.~Prabhu and M.~M.~Roberts,
  arXiv:1609.06729 [hep-th].

\bibitem{Geracie:2016bkg}
  M.~Geracie,
  arXiv:1611.01198 [hep-th].


\bibitem{LevyLeblond:1967zz}
  J.~M.~Levy-Leblond,
  Commun.\ Math.\ Phys.\  {\bf 6} (1967) 286.
  doi:10.1007/BF01646020

\bibitem{Fuini:2015yva}
  J.~F.~Fuini, A.~Karch and C.~F.~Uhlemann,
  Phys.\ Rev.\ D {\bf 92} (2015) no.12,  125036
  doi:10.1103/PhysRevD.92.125036
  [arXiv:1510.03852 [hep-th]].


\bibitem{Duval:1995fa}
  C.~Duval, P.~A.~Horvathy and L.~Palla,
  Annals Phys.\  {\bf 249} (1996) 265
  doi:10.1006/aphy.1996.0071
  [hep-th/9510114].


\bibitem{Christensen:1978md}
  S.~M.~Christensen and M.~J.~Duff,
  Nucl.\ Phys.\ B {\bf 154} (1979) 301.

\bibitem{Freedman}
"Supergravity", Daniel Z. Freedman and Antoine van Proyen (CUP, 2012).


\bibitem{Solodukhin:2009sk}
  S.~N.~Solodukhin,
  JHEP {\bf 1004} (2010) 101
  doi:10.1007/JHEP04(2010)101
  [arXiv:0909.0277 [hep-th]].
  
\bibitem{Arav:2017plg}
  I.~Arav, I.~Hason and Y.~Oz,
  arXiv:1702.00690 [hep-th].

 \bibitem{Anselmi:1997am}
  D.~Anselmi, D.~Z.~Freedman, M.~T.~Grisaru and A.~A.~Johansen,
  Nucl.\ Phys.\ B {\bf 526} (1998) 543
  [hep-th/9708042].
  
\bibitem{Auzzi:2015yia}
  R.~Auzzi and B.~Keren-Zur,
  JHEP {\bf 1505} (2015) 150
  doi:10.1007/JHEP05(2015)150
  [arXiv:1502.05962 [hep-th]].

\bibitem{Andringa:2013mma}
  R.~Andringa, E.~A.~Bergshoeff, J.~Rosseel and E.~Sezgin,
  Class.\ Quant.\ Grav.\  {\bf 30} (2013) 205005
  doi:10.1088/0264-9381/30/20/205005
  [arXiv:1305.6737 [hep-th]].





\end{thebibliography}
\end{document}